\documentclass[12pt,aps]{revtex4}
\usepackage[cp1251]{inputenc}
\input{epsf}
\begin{document}
\author{E.B.Aleksandrov, V.S.Zapasskii, G.G.Kozlov, A.C.Selden}
\title{Open Letter to the OSA Topical Meeting "Slow and Fast Light"}

\vskip10mm
\begin{abstract}
 The question is raised about whether the pulse delay in a saturable absorber
has anything to do with the "slow light" and group velocity reduction. We
also wonder why the 40-years-old trivial interpretation of the effect remains
so far ignored.
\end{abstract} 
\maketitle \vskip10mm

 Dear Colleagues:

One of the approaches to the slow light technique is based on the effect of coherent
 population oscillations (CPO). First studies of the CPO-based slow light were performed
 on ruby crystal by Bigelow et al \cite{1}.

        We would like to point out  again that the effect of pulse delay in a saturable
         absorber  has been known for more
         than 40 years (see, e.g., \cite{2,3,4,5}), with all
its manifestations (both in time and frequency domains)
being perfectly interpreted within the model of rate equations and absorption
dynamics. Nowadays, 40 years later, this interpretation, based on a simplest
nonlinear optical effect, seems
fairly trivial.  Still, at the beginning of our century, this effect was
revised and attributed to group velocity reduction due to steep
 dispersion at the CPO-related spectral hole.

Our disagreement with this revision was presented
in publications \cite{6,7,8} and in the invited talk
         "Slow Light: Underlying Controversies" given at the
previous SL Topical Meeting.
 Still, common agreement on the point is not so far reached.
 The questions that remain unanswered and could clarify the
situation are:

(i) What new has been found in the experiments  \cite{1}
 compared with what has been  known since 1967?

(ii) What is wrong in the old model of pulse delay in
a saturable absorber\cite{2,3,4,5}? Or perhaps, in reality, "slow light"
has been discovered 40 years ago?

(iii) How to distinguish between the CPO-based slow light and the trivial effect of
pulse delay in a saturable absorber?

(iv) Why in the papers on CPO-based slow light this trivial interpretation of the effect is not mentioned and no
references to the above old publications \cite{2,3,4,5} are given?

In our opinion the SL  Topical Meeting is the appropriate forum for an open discussion of the
issues we have raised, with the aim of achieving a definitive  resolution of this
problem.

\end{document}